\documentclass[aps,prl,twocolumn,superscriptaddress,10pt]{revtex4-1}

\usepackage{graphicx}

\usepackage{amsmath}

\usepackage{color}

\usepackage{epstopdf}

\usepackage{multirow,bigstrut}

\usepackage{times,mathptmx,helvet}

\DeclareSymbolFont{newfont}{OML}{cmm}{m}{it}
\DeclareMathSymbol{\Epsilon}{3}{newfont}{15}

\usepackage[
pdfstartview=XYZ,
bookmarks=false,
colorlinks=true,
linkcolor=blue,
urlcolor=blue,
citecolor=blue,
pdftex
]{hyperref}

\makeatletter

\newcommand{\Rmnum}[1]{\expandafter\@slowromancap\romannumeral #1@}
\makeatother

\begin{document}


\title{\large{Hot Spot Evolution Measured by High-Resolution X-Ray Spectroscopy \\
at the National Ignition Facility}\vspace{3pt}}

\author{Lan~Gao}
\affiliation{Princeton Plasma Physics Laboratory, Princeton University, Princeton, NJ, 08543, USA} 

\author{B. F. Kraus}
\affiliation{Princeton Plasma Physics Laboratory, Princeton University, Princeton, NJ, 08543, USA} 

\author{K. W. Hill}
\affiliation{Princeton Plasma Physics Laboratory, Princeton University, Princeton, NJ, 08543, USA} 

\author{M. B. Schneider}
\affiliation{Lawrence Livermore National Laboratory, Livermore, California 94550, USA} 

\author{A. Christopherson}
\affiliation{Laboratory for Laser Energetics, University of Rochester, Rochester, NY, 14623, USA}

\author{B. Bachmann}
\affiliation{Lawrence Livermore National Laboratory, Livermore, California 94550, USA}

\author{M. Bitter}
\affiliation{Princeton Plasma Physics Laboratory, Princeton University, Princeton, NJ, 08543, USA} 

\author{P. Efthimion}
\affiliation{Princeton Plasma Physics Laboratory, Princeton University, Princeton, NJ, 08543, USA}

\author{N. Pablant}
\affiliation{Princeton Plasma Physics Laboratory, Princeton University, Princeton, NJ, 08543, USA}

\author{R. Betti}
\affiliation{Laboratory for Laser Energetics, University of Rochester, Rochester, NY, 14623, USA} 

\author{C. Thomas}
\affiliation{Laboratory for Laser Energetics, University of Rochester, Rochester, NY, 14623, USA} 

\author{D. Thorn}
\affiliation{Lawrence Livermore National Laboratory, Livermore, California 94550, USA} 

\author{A. G. MacPhee}
\affiliation{Lawrence Livermore National Laboratory, Livermore, California 94550, USA} 

\author{S. Khan}
\affiliation{Lawrence Livermore National Laboratory, Livermore, California 94550, USA} 

\author{R. Kauffman}
\affiliation{Lawrence Livermore National Laboratory, Livermore, California 94550, USA} 

\author{D. Liedahl}
\affiliation{Lawrence Livermore National Laboratory, Livermore, California 94550, USA} 

\author{H. Chen}
\affiliation{Lawrence Livermore National Laboratory, Livermore, California 94550, USA} 

\author{D. Bradley}
\affiliation{Lawrence Livermore National Laboratory, Livermore, California 94550, USA}

\author{J. Kilkenny}
\affiliation{Lawrence Livermore National Laboratory, Livermore, California 94550, USA}

\author{B. Lahmann}
\affiliation{Massachusetts Institute of Technology, Cambridge, MA 02139, USA}

\author{E. Stambulchik} 
\affiliation{Faculty of Physics, Weizmann Institute of Science, Rehovot 7610001, Israel}

\author{Y. Maron}
\affiliation{Faculty of Physics, Weizmann Institute of Science, Rehovot 7610001, Israel}

\date{\today}

\vspace{4 cm}

\begin{abstract}


Evolution of the hot spot plasma conditions was measured using high-resolution x-ray spectroscopy at the National Ignition Facility (NIF). The capsules were filled with DD gas with trace levels of Kr, and had either a high-density-carbon (HDC) ablator or a tungsten (W)-doped HDC ablator. Time-resolved measurement of the Kr He$\beta$ spectra, absolutely calibrated by a simultaneous time-integrated measurement, allows inference of the electron density and temperature through observing Stark broadening and the relative intensities of dielectronic satellites. By matching the calculated hot spot emission using a collisional-radiative code to experimental observations, the hot spot size and areal density are determined. These advanced spectroscopy techniques further reveal the effect of W dopant in the ablator on the hot spot parameters for their improved implosion performance. 

\end{abstract}


\maketitle

In inertial confinement fusion (ICF), a powerful driver is used to rapidly compress the fuel to a stagnated plasma in a central hot spot with fusion-relevant density and temperature conditions \cite{lindl1995}. The stagnated state evolves dramatically over time in real experiments, making accurate time-resolved measurements of the plasma parameters in the hot spot of crucial importance for assessing the implosion performance and achieving ignition \cite{Miller_2004, Moses2009}.

In the majority of current ICF experiments, the electron temperature $T_{\rm e}$ is inferred from the slope of the Bremsstrahlung spectrum emitted from the hot spot \cite{Jarrott_PRL_2018}. The ion temperature $T_{\rm i} $ is measured from neutron time-of-flight (nTOF) spectrometers \cite{Glebov_RSI_2010} and the density $n_{\rm e}$ is deduced from $T_{\rm i} $ assuming uniform plasma conditions across the hot spot. Besides being spatially and temporally integrated, the neutron spectral measurements are affected by plasma flows and implosion asymmetry resulting in ambiguity in the data inference \cite{Maria_PRE_2016}. 

High-resolution x-ray spectroscopy has been actively investigated at ICF facilities to expand the measurement capabilities. By adding low- (Neon) or middle-Z (Argon) elements in the capsule fuel and studying their characteristic line emissions, important information such as hot spot plasma parameters \cite{Yaakobi_PRL_1977,Hammel_PRL_1993,Woolsey_PRE_1997,Regan_PoP_2002, Florido_PRE_2011} and their spatial profiles \cite{Golovkin_2002, Welser2007} were determined. By embedding mid-Z dopant layers in the capsule ablator and studying mix of the dopants into the hot spot, hot spot asymmetries and mix were obtained \cite{Regan_2013_PRL}.

A high-Z dopant material is required to approach and diagnose the hot spot created by implosions of the NIF scale, such that their x-ray photons can propagate through the remainder of the fuel as well as the dense compressed shell without significant attenuation. To that end, a capsule with Kr-doped fuel was developed utilizing the 13-16 keV Kr K-emission lines from implosion cores \cite{Ma_RSI_2016}. A time-resolved, high-resolution x-ray Bragg crystal spectrometer named dHIRES (DIM-based high-resolution spectrometer) was developed to capture the Kr emissions with a high resolving power ($E/\Delta E$ $\sim$ 1000--2000) \cite{Hill_RSI_2016}. In addition to record the Kr He$\alpha$ + Ly$\alpha$ and He$\beta$ complexes onto a streak camera, a simultaneous time-integrated measurement provides {\it in-situ} calibrations for the streak camera signals. The dHIRES was absolutely calibrated, with its measured integrated reflectivity, energy dispersion, and energy resolution playing a decisive role in ensuring the success of data acquisition and analyses for all channels \cite{Gao_RSI_2018}.

This letter reports the first time-resolved simultaneous measurements of the hot spot plasma density, temperature, size and areal density for Kr-doped capsule implosions at their stagnation phase using high-resolution x-ray spectroscopy at the NIF. The capsules had either a high-density-carbon (HDC) ablator or a tungsten (W)-doped HDC ablator enclosing Kr-doped DD gas. Time-resolved Kr He$\beta$ spectra measured by dHIRES allow inference of the electron density and temperature through observing Stark broadening and the relative intensities of dielectronic satellites. The high spectral resolution enables detailed line-shape study and comparison to Stark broadening calculations, here fully treating ion dynamics effects. 
By matching the calculated hot spot emission using a collisional-radiative code to experimental observations, the hot spot size is determined, which then allows inference of the hot spot areal density. Compared to the undoped HDC implosion, the W-doped HDC capsule is shown to have higher hot spot density, temperature and areal density as well as smaller hot spot size, providing direct evidences of increased stagnation pressure and thermonuclear performance by using a high-Z dopant in the ablator. These measurements, showing temporal profiles of the core plasma parameters with high precision, are essential to benchmark numerical simulations aiming for ignition.

\begin{figure}[t]
\includegraphics{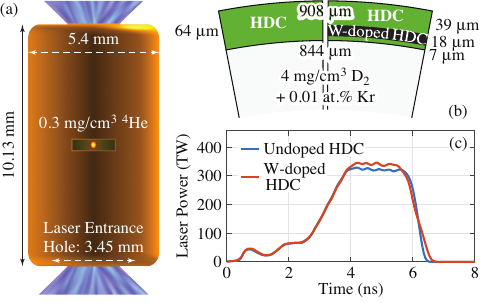}
\caption{\label{fig0} (color online). (a) Hohlraum parameter. The rectangle indicates the viewing window for dHIRES. The yellow dot indicates the capsule. (b) Pie diagrams of the undoped HDC capsule and W-doped HDC capsule. (c) Laser pulses delivered to the capsules.}
\end{figure}

The experiments described herein used the existing BigFoot platform to create a well-behaved hot spot \cite{Baker_PRL_2018,Casey2018}. The BigFoot approach aims at good control of hohlraum symmetry and hydrodynamic instabilities with short pulse shapes and at high adiabat, thus improving the implosion predictability and performance. The target is a symmetric capsule (symcap), which is an ignition capsule surrogate that has the appropriate ablator thickness and fuel gas to achieve symmetric implosions \cite{Kyrala_2011}. Figure \ref{fig0} illustrates the experimental details. 192 NIF laser beams heat a Au hohlraum generating a radiation bath to implode the capsule suspended at the center of the hohlraum. Two campaigns were performed at very similar laser conditions. Both capsules had the same dimensions and were filled with Kr-doped DD gas at 0.01 atomic percent (at.$\%$). The Kr fielding temperature was 80 K maintaining the gaseous phase for both Kr and DD. One capsule had an ablator shell made of HDC, refereed to as the undoped HDC capsule, while the other had 0.21 at.$\%$ of W in a 18-$\mu$m-thick layer, referred to as the W-doped HDC capsule. Previous NIF experiments using the NIF X-ray spectrometer (NXS), while not being able to provide measurements on $n_{\rm e}$ and $T_{\rm e}$ due to the low resolving power ($E/\Delta E$ $\ge$ 60), demonstrated the feasibility of using Kr to extract x-ray spectroscopic information emitted from the hot spot \cite{Chen_PoP_2017}. The 0.01 at.$\%$ of Kr was shown to have a minimal perturbation on the implosion \cite{Ma_RSI_2016}.

The dHIRES was fielded at an equatorial line-of-sight at (90, 315) to acquire a clean Kr emission without contributions from the Au hohlraum \cite{Ma_RSI_2016}. There is a viewing window cut out of the hohlraum wall and replaced with a 80-$\mu$m-thick HDC foil along the signal line-of-sight to contain the hohlraum plasma and transmit the x-ray emission from the capsule. The window is 1.7 mm $\times$ 0.5 mm, large enough to cover the entire x-rays diffracted from the crystals. The Kr He$\alpha$ + Ly$\alpha$, and He$\beta$ complexes are focused by two conical crystals onto a streak camera with a temporal resolution of $\sim$25 ps \cite{Kimbrough_RSI_2001}. A third cylindrical crystal in the von H\'amos configuration diffracts x-rays of the entire energy range onto an image plate providing a simultaneous time-integrated spectral measurement. The analyses here focus on the He$\beta$ spectra that are optically thin allowing inference of hot spot plasma parameters. 

\begin{figure}[t]
\includegraphics{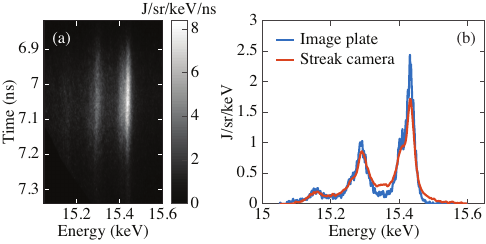}
\caption{\label{fig1} (color online). (a) The cross-calibrated streak camera data for the He$\beta$ complexes from undoped HDC experiment N171102. (b) The time-integrated streak camera data is overlaid on top of the simultaneous image plate measurement.}
\end{figure}

Figure \ref{fig1}(a) shows the cross-calibrated streak camera data for the He$\beta$ complexes from the undoped HDC experiment N171102. The entire x-ray emission at stagnation was captured showing an overall duration of Kr emission of $\sim$300 ps. Pre-shot simulations with rad-hydro code LASNEX \cite{Zimmerman_1975} post-processed with the collisional-radiative code Cretin \cite{SCOTT_2001}, combined with the measured spectrometer throughput, allowed an optimal selection of filtering and therefore signal counts on the camera. Details of the He$\beta$ resonance and satellite lines are clearly seen; this is contrast to previous measurements with NXS in which these separate peaks could not be resolved. By integrating the streak camera data in both time and energy and accounting for the filter transmission, the total signal count represents the source emission in the Kr He$\beta$ region, which is equal to that obtained from the simultaneous image plate measurement taken the image plate channel filter transmission into account and integrated over the same spectral energy range. This process provides the conversion factor of the streak camera data from counts per pixel area to a physics value in J/sr/keV/ns as shown in Figure \ref{fig1}(a). The vertical axis is time measured on shot. The horizontal axis represents x-ray energy calculated from the calibrated spectral dispersion \cite{Gao_RSI_2018}. Figure \ref{fig1}(b) shows the time-integrated x-ray spectrum from streak camera overlaid on top of the image plate measurement for the same energy of interest confirming the validity of the calibration process. The streak camera spectrum has a higher valley between the He$\beta$ resonance line and the Li-like satellite line and a tall tail on the blue side of the resonance line. This is due to its lower resolving power as well as small contributions from the space-charge effect among the streak camera pixels. X-ray continuum emissions were subtracted and only the Kr line emissions were used in the cross-calibration process in Figure \ref{fig1}(b). Another independent cross-calibration process including both continuum and line emissions provided a same conversion factor for the streak camera data confirming no contribution from the Au emission was observed by dHIRES. 

\begin{figure}[t]
\includegraphics{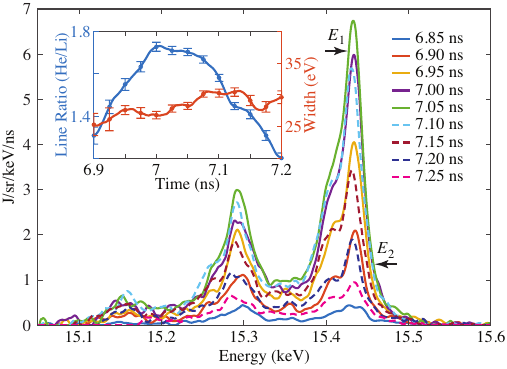}
\caption{\label{fig2} (color online). History of the high-resolution Kr He$\beta$ spectra for N171102. {\it E}$_1$ and {\it E}$_2$ denotes the energy thresholds at which the widths of the Stark broadened Kr He$\beta$ complexes were determined for the spectrum at 7.05 ns (green). The inset shows the derived He/Li line ratios (blue) and the Stark widths (red).}
\end{figure}

Figure \ref{fig2} shows the time-resolved Kr He$\beta$ spectra every 50 ps directly deduced from the calibrated streak camera image. Each spectrum represents the averaged signal from 48 consecutive spectral lineouts of the streak camera image to account for the 25-ps temporal resolution. The peak Kr emission is seen at 7.05 ns, consistent with the x-ray bang time (BT) measured from SPIDER (Streaked Polar Instrumentation for Diagnosing Energetic Radiation) \cite{Khan_2012}. The x-ray BT is the time when peak x-ray emission from the implosion core occurs. The central peak of the He$\beta$ resonance line near 15.43 keV continues to shift to lower energies until $\sim$ 7.15 ns and then to higher energies. The width of the resonance line and its line intensity with respect to the dielectronically excited Li-like line near 15.29 keV also vary in time. All are clear signatures of the evolving plasma conditions during this phase. 

\begin{table*}[t]
\begin{center}
\caption{Summary of the undoped and W-doped HDC experiments.}
\label{ExpSummary}
\tabcolsep=0.18cm
\begin{tabular}{c|cc|ccccccc}
\hline\hline
 & Laser & Laser & X-ray BT & X-ray BT & DD Yield & Ti & $\rm \rho R_{hs}$  & Te & Hot-spot P0 \\
Ablator & energy & power & SPIDER & dHIRES & nTOF & nTOF & nTOF & Continuum & Penumbral imaging \\
\cline{2-10}
& MJ & TW & ns & ns & 10$^{13}$ & keV & $\rm mg/cm^2$ & keV & $\mu$m \\
\hline\hline
Undoped HDC & 1.08 & 327 & 7.05 & 7.05 & 1.0$\,\pm\,$0.04 & 3.27$\,\pm\,$0.16 & 44.4$\,\pm\,$1.92 &  2.56$\,\pm\,$0.25 & 73.5$\,\pm\,$0.5 \\
\hline
W-doped HDC & 1.11 & 342 & 7.16 & 7.16 & 2.0$\,\pm\,$0.08 & 3.68$\,\pm\,$0.16 & 65.2$\,\pm\,$2.56  & 2.90$\,\pm\,$0.25 & 60.0$\,\pm\,$0.5  \\
\hline\hline
\end{tabular}
\vspace{-3mm}
\end{center}
\vspace{-5mm}
\end{table*}

At our estimated hot spot plasma conditions, the Stark broadening of the Kr He$\beta$ complex largely depends on the plasma density \cite{STAMBULCHIK2006730}, while the intensity ratio between the He$\beta$ complex and the Li-like Kr line is sensitive to the electron temperature \cite{Chen_PoP_2017}. The inset of Figure \ref{fig2} presents the experimentally measured He/Li line ratio and Stark width of the He$\beta$ complex as a function of time calculated from the time-resolved He$\beta$ spectra. The He/Li line ratio is determined by the total line intensity of the He$\beta$ complex from 15.37 keV to 15.5 keV relative to that from the Li-like 15.23--15.37 keV satellite features. With time, the He/Li line ratio increases to a peaked value at $\sim$7 ns followed by a decay until the end of the emission. Accurately identifying the Stark broadening effect is more complicated. The He$\beta$ complex is comprised of three components: a strong central peak at $\sim$15.43 keV and two relatively weaker peaks at its low and high energy sides. A Li-like spectral line is underneath the low energy peak contributing to the entire low-energy features in the vicinity of 15.42 keV. To avoid the uncertainties in line broadening from the Li-like satellite line, the Stark width is determined as the energy range between {\it E}$_1$ and {\it E}$_2$, where {\it E}$_1$ is the energy corresponding to 90$\%$ of the central peak height and at {\it E}$_2$ the spectral height falls to 20$\%$ of the peak height. A slow increase in the measured Stark width is seen until $\sim$7.13 ns. The error bars associated with each line ratio and width include statistical errors directly from the measured counts and $\sim$5$\%$ systematic errors primarily from uncertainties in calibrating the streak camera data.

\begin{figure}[t]
\includegraphics{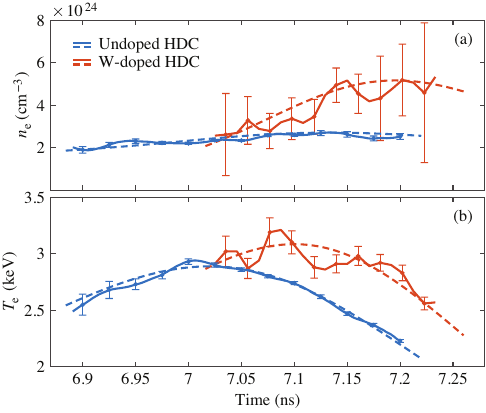}
\caption{\label{fig3} (color online). The experimentally measured $n_{\rm e}$ and $T_{\rm e}$ (solid lines) for the undoped and W-doped HDC capsules compared with the spatially averaged, x-ray weighted calculations (dashed lines).}
\end{figure}

Adding a high-Z dopant in the ablator shell has been proposed to improve the implosion performance \cite{Hopkins_2018}. Advantages include reduction of the capsule preheat from the Au M-band x-ray radiation and therefore increased capsule compression and enhanced stability at the fuel-capsule interface. The experiment reported here for the W-doped HDC target (N180109) directly links to the first role as there is no DT fuel layer in the capsule. The W-doped target was driven at almost same laser conditions and probed with the same diagnostics as the undoped experiment. The x-ray bang time was measured at $\sim$7.16 ns. Using the same data analysis processes, both image plate and streaked channels show more broadened He$\beta$ resonance complex and larger He/Li line intensity ratios than those from the undoped target, indicating the effect of the W dopant on hot spot plasma parameters.

Figure \ref{fig3} presents the inferred hot spot $n_{\rm e}$ and $T_{\rm e}$ as a function of time for both experiments, by comparing the experimentally measured Stark widths and He/Li line ratios to theoretical calculations. All calculations were convolved with 12-eV Gaussian functions to correct for the instrumental response. With increasing $n_{\rm e}$, the amplitudes of the low and high energy He$\beta$ peaks relative to that of the central peak increase causing the entire complex broadened. By matching the experimentally measured line width to detailed line-shape calculations that include full ion dynamics effects \cite{STAMBULCHIK2006730} at the same energy thresholds {\it E}$_1$ and {\it E}$_2$, $n_{\rm e}$ was inferred at each measurement time. For the undoped HDC experiment, $n_{\rm e}$ slowly grows from 2 $\times$ 10$^{24}$ cm$^{-3}$ at 6.9 ns to $\sim$2.6 $\times$ 10$^{24}$ cm$^{-3}$ at 7.13 ns, $\sim$80 ps after the x-ray bang time. Details of the line-shape studies and their comparison to our measured spectral profiles for density inference are prepared in another paper \cite{Hill2022}. To infer $T_{\rm e}$, calculations with a grid of $n_{\rm e}$-$T_{\rm e}$ conditions were performed using the collisional-radiative atomic code SCRAM \cite{HANSEN_2007}, generating He/Li line ratios as functions of $n_{\rm e}$ and $T_{\rm e}$. Using $n_{\rm e}$ from Stark broadening and the experimentally measured He/Li line ratios shown in Figure \ref{fig2}, $T_{\rm e}$ was inferred. For the undoped experiment, $T_{\rm e}$ increases from $\sim$2.55 keV to $\sim$2.93 keV at $\sim$7 ns, and then slowly decreases to $\sim$2.23 keV at the end of the measurement.

\begin{figure}[b]
\includegraphics{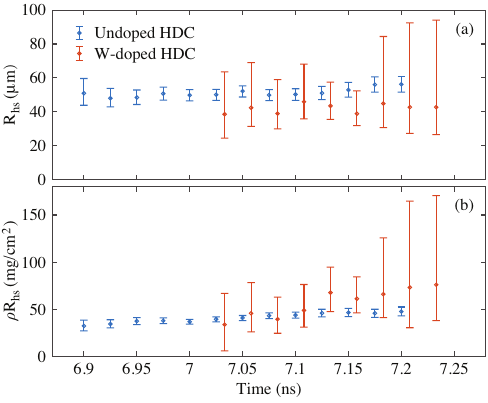}
\caption{\label{fig4} (color online). Evolution of the hot spot radius $\rm R_{hs}$ and areal density $\rho\rm R_{hs}$ for the undoped and W-doped HDC shots.}
\end{figure}

Low signal levels obtained from the W-doped HDC experiment N180109 due to purposely chosen thick filters for the streaked channels lead to large statistical errors and fluctuations in the analyzed data. Nonetheless several clear features are presented comparing the two experiments. First, both experiments exhibit the same evolution trend for $n_{\rm e}$ and $T_{\rm e}$, where $T_{\rm e}$ peaks $\sim$50-60 ps before the x-ray bang time and $n_{\rm e}$ continues to grow afterwards.  Secondly, compared to the  undoped target, both $n_{\rm e}$ and $T_{\rm e}$ are increased for the W-doped target, with peak $n_{\rm e}$ approaching $\sim$5.2 $\times$ 10$^{24}$ cm$^{-3}$ and peak $T_{\rm e}$ approaching $\sim$3.21 keV. This leads to an increase in the stagnation pressure for N180109 (peaked at $\sim$48 Gbar) comparing to N171102 (peaked at $\sim$24 Gbar). The nTOF measurement further supports these analyses. An increase in neutron yield, fuel $\rho\rm R_{hs}$ and DD neutron burn averaged $T_{\rm i}$ was measured for the W-doped HDC target, as shown in Table \ref{ExpSummary} that summarizes the performance metrics for both shots. Our results therefore show strong evidence of increased stagnation pressure as a result of increased both density and temperature and enhanced thermonuclear output by using W dopant in the ablator.

It is important to understand that the time-dependent $n_{\rm e}$ and $T_{\rm e}$ measured by dHIRES are spatially averaged over the entire hot spot and x-ray weighted. Hot spot spatial profiles can be directly measured using monochromatic x-ray line images of the entire core and novel spectroscopic analyses \cite{Golovkin_2002, Welser2007}. The BigFoot HDC symcap implosions used in dHIRES measurement were well modeled, controlled and reproducible. It is sufficiently accurate to use the isobaric model by Betti {$\textit{et al.}$} \cite{Betti_2001} to calculate the spatiotemporal profiles of the hot spot $n_{\rm e}$ and $T_{\rm e}$, with their central values calculated using a 1-D thick-shell model \cite{Christopherson_2018}. An emissivity database was generated for a grid of $n_{\rm e}$ and $T_{\rm e}$ using SCRAM. The spatially averaged, x-ray weighted $n_{\rm e}$({\it t}) and $T_{\rm e}$({\it t}) were then calculated using the radial profiles of the hot spot and the associated x-ray emissivity at each position of the hot spot along the measurement line of sight. Here $t$ represents time. Figure \ref{fig3} shows that the simulated $n_{\rm e}$({\it t}) and $T_{\rm e}$({\it t}) are benchmarked to the experimental measurements.

The absolutely calibrated Kr He$\beta$ spectra allow accurate determination of the hot spot size. This is because the cold shell can not excite such high-energy x rays and dHIRES measurement are purely from the Kr homogeneously mixed in the DD fuel. Taking the spatiotemporal profiles of the hot spot $n_{\rm e}$ and $T_{\rm e}$ that match the experimental observations, the hot spot size is determined by matching the measured Kr He$\beta$ signals in simulations (Figure \ref{fig4}(a)). Details of the spectral comparison and error bar analysis for the inferred hot spot size are shown in the Supplemental Material \cite{Supp1}. Despite large error bars for the W-doped HDC experiment, their hot spot sizes are consistently smaller than those from the undoped HDC shot. The x-ray weighted, time-averaged hot spot size was also calculated, 51.5$\,\pm\,$4.0 $\mu$m and 40.6$\,\pm\,$20.1 $\mu$m for the undoped and W-doped HDC experiment respectively, significantly smaller than those measured from the penumbral x-ray imaging technique \cite{Bachmann_PRE_2020}.

Combining the experimentally measured $n_{\rm e}$({\it t}) and $\rm R_{hs}(\it t)$, hot spot areal density $\rm \rho R_{hs}({\it t)}$ was calculated assuming equal ion density to electron density. As shown in Figure \ref{fig4}(b), evolution of $\rm \rho R_{hs}$ predominantly follows $n_{\rm e}$({\it t}) since the measured $\rm R_{hs}(\it t)$ remains about constant which is expected at the stagnation phase of the implosions. Relative to the undoped HDC experiment, $\rm \rho R_{hs}({\it t})$ for the W-doped HDC shot is consistently larger with the inferred increase in stagnation pressure. The x-ray weighted, time-averaged hot spot $\rm \rho R_{hs}$ was also calculated, significantly lower than those measured from nTOF (34.4$\,\pm\,$3.95 $\rm mg/cm^2$ vs 44.4$\,\pm\,$1.92 $\rm mg/cm^2$ for the undoped experiment and 45.3$\,\pm\,$28.42 vs 65.2$\,\pm\,$2.56 $\rm mg/cm^2$ for the W-doped experiment). Inference of $\rm \rho R_{hs}$ from nTOF relies on accurate evaluation of $T_{\rm e}$ in their modeling and could be affected by material mixing in the fuel~\citep{Rinderknecht_PoP_2015}.

In summary, evolution of the hot spot was measured for Kr-doped capsule implosions at the NIF, made possible by detailed line-shape studies from high-resolution x-ray spectroscopy. The NIF dHIRES provides time-resolved, absolutely calibrated Kr He$\beta$ spectra allowing simultaneous inference of the hot spot electron density, temperature, size and areal density. The time-averaged hot spot size and areal density based on time-resolved dHIRES measurements point to overestimation of these critical parameters by traditional x-ray imaging and nTOF. The experiments that were executed, with and without W dopant in the HDC shell, show consistent evolution trend for the electron density and temperature. Our measurements also provide direct evidence that using a high-Z dopant in the ablator increases the stagnation pressure and enhances thermonuclear performance, as a result of increased hot spot electron density and temperature. Precision measurements of the core plasma parameters with high temporal resolutions, as shown in our technique, will be of crucial importance for benchmarking numerical simulations aiming for ignition. 

\begin{acknowledgments}
This work was performed under the auspices of the U.S. Department of Energy by Princeton Plasma Physics Laboratory under contract DE-AC02-09CH11466 and by Lawrence Livermore National Laboratory under contract DE-AC52-07NA27344. L. G., B. F. K., K. W. H. and P. E. would like to thank Prof. R. Mancini for useful discussions.
\end{acknowledgments}

\bibliography{1MAIN}

\end{document}